\documentclass[twocolumn,showpacs,preprintnumbers,amsmath,amssymb]{revtex4}

\usepackage{graphicx}
\usepackage{dcolumn}
\usepackage{bm}

\begin{document}


\title{Synchronized Collective Behavior via Low-cost Communication}

\author{Hai-Tao Zhang$^1$, Michael ZhiQiang Chen$^{1,\dagger}$, and Tao Zhou$^{2,3}$}
\affiliation{$^1$Department of Engineering,
University of Cambridge, Cambridge CB2 1PZ, U.K. \\
$^2$Department of Modern Physics, University of Science and
Technology of China, Hefei 230026, PR China\\
$^3$Department of Physics, University of Fribourg, Chemin du Muse,
CH-1700 Fribourg, Switzerland}

\begin{abstract}
An important natural phenomenon surfaces that satisfactory
synchronization of self-driven particles can be achieved via sharply
reduced communication cost, especially for high density particle
groups with low external noise. Statistical numerical evidence
illustrates that a highly efficient manner is to distribute the
communication messages as evenly as possible along the whole dynamic
process, since it minimizes the communication redundancy. More
surprisingly, it is discovered that there exist some abnormal
regions where moderately decreasing the communication cost can even
improve the synchronization performance. A phase diagram on the
noise-density parameter space is given, where the dynamical
behaviors can be divided into three qualitatively different phases:
{\it normal phase} where better synchronization corresponds to
higher communication cost, {\it abnormal phase} where moderately
decreasing communication cost could even improve the
synchronization, and the {\it disordered phase} where no coherence
among individuals is observed.
\end{abstract}


 \maketitle

Over the last decade or so, physicists have been looking for common,
possibly universal, features of the collective behaviors of animals,
bacteria, cells, molecular motors, as well as driven granular
objects. Thus, the collective motion of a group of autonomous
particles is a subject of intensive research that has potential
applications in biology, physics and engineering. One of the most
remarkable characteristics of systems such as a flock of birds, a
school of fish, or a swarm of locusts is the emergence of states of
collective order in which the particles move in the same direction,
i.e. {\it ordered state} \cite{vi95,gr04,al07}, despite the fact
that the interactions are (presumably) of short range. Moreover,
this ordered state seeking problem can be further generalized to a
consensus problem \cite{sa04}, i.e. groups of self-propelled
particles agreeing upon certain quantities of interest like
attitude, position, temperature, voltage and so on. Distributed
computation based on solving consensus problems has direct
implications on sensor network data fusion, load balancing,
swarms/flocks, unmanned air vehicles (UAVs), attitude alignment
satellite clusters, congestion control of communication networks,
multi-agent formation control  and so on \cite{ak02,og04,ar02}.

In Ref. \cite{vi95}, a dynamical model describing the collective
motion is proposed in a system of self-propelled particles. Due to
its simplicity yet efficiency, this so-called Vicsek model has been
drawing more and more attention recently and gaining increased
popularity from both physics and engineering communities
\cite{he00,ch06,co03,co05,do06,ga03,gr04,al07,mo04,he05}. In the
Vicsek model each particle tends to move in the average direction of
motions of its neighbors while being simultaneously subjected to
noise. As the amplitude of the noise increases the system undergoes
a phase transition from an \emph{ordered state} in which the
particles move in the same direction, to a \emph{disordered state}
in which the particles move independently in random directions.
Gr\'{e}goire and Chat\'{e} \cite{gr04} modified the Vicsek model by
changing the way in which the noise is introduced into the group. By
this means, the phase transition is switched from second to first
order. More recently, in order to stabilize flocks/swarms,
Gazi-Passino \cite{ga03} and Moreau \cite{mo04} developed two
alternative models, i.e. the Attraction/Repulsion (A/R) model and
the linearized model, respectively. The former yields a cohesive
swarm with bounded size in a finite time, while the latter can
guarantee the convergence of all the particles' states to a common
one with {\it complete communication}, i.e. sending messages all
along.

In brief, based on complete communication, most of the previous
models of self-propelled particle groups yield many attractive
characteristics like convergence, ordered state, consensus,
rendezvous, cohesion, robustness, etc. However, in this Letter, an
important phenomenon is discovered that complete communication is
not the most efficient manner. For many kinds of self-propelled
particle groups and natural swarms/flocks/schools, satisfactory
ordered state performances can still be achieved with sharply
reduced communication cost. Secondly, even more surprisingly, there
exist some abnormal regions in the density-noise space where
moderately reducing the communication cost can help increase the
performance. A general physical picture behind our finding is as
follows: in abundant natural bio-groups composed of animals,
bacteria, cells and so on, each particle does not send messages
throughout the whole process, but now and then in some suitable
manner, which is called {\it partial communication}. Some close
examples can be found in firefly groups, deep-sea luminous fish
schools and so on. Each particle uses light signal with limited
power to guide the others, and just flashes at some suitable
discrete times to save energy, which yields satisfactory collective
performances. Other than the above mentioned natural phenomena, our
work is also partially inspired by Ref. \cite{co05}, in which it is
revealed that the larger the group the smaller the proportion of
informed individuals needed to guide the whole group, and that only
a very small proportion of informed individuals is required to
achieve great accuracy. In addition, we found the role of
information redundancy on the present model: The higher the
redundancy, the worse the synchronization performance. Therefore, a
highly efficient manner is to distribute the communication messages
as evenly as possible along the whole dynamic process. From an
industrial application point of view, the phenomena and strategies
reported in this Letter may be applicable in some relevant
prevailing engineering areas like autonomous robot formations,
sensor networks, UAVs and so on. Since each particle in these groups
has just limited power to send messages, partial communication is
required to save energy \cite{ak02,og04,ar02}.

Due to its popularity, we will focus our simulation and
investigation on the Vicsek model \cite{vi95}. In this model, the
velocities $\{{\bf v}_i\}$ of $N$ particles are determined
simultaneously at each time step, and the position of the $i$th
particle is updated according to ${\bf x}_i (t+1)={\bf x}_i(t)+{\bf
v}_i(t)\Delta t$. Here the velocity of a particle ${\bf v}_i (t+1)$
is constructed to have an absolute value $v$ and a direction given
by the angle $\theta (t+1)$. This angle is obtained from the
expression
\begin{equation} \label{Eq Vicsek model angle}
\theta (t+1)=\left<\theta\left(t\right)\right>_r+\Delta\theta
\end{equation}
where $\left<\theta\left(t\right)\right>_r$ denotes the average
direction of the velocities of particles (including particle $i$)
being within a circle of radius $r$ surrounding the given particle
$i$. The average direction is given by
\begin{equation}
\label{average direction}
\left<\theta\left(t\right)\right>_r=\mbox{arctan}\left[\left<\mbox{sin}\left(\theta\left(t\right)\right)\right>_r/
\left<\mbox{cos}\left(\theta\left(t\right)\right)\right>_r\right]
\end{equation}
where $\left<\mbox{sin}\left(\theta\left(t\right)\right)\right>_r$
and $\left<\mbox{cos}\left(\theta\left(t\right)\right)\right>_r$
denote the average sine and cosine values of the velocities
respectively, and $\Delta\theta$ represents a random noise obeying a
uniform distribution in the interval $\left[-\eta/2, \eta/2\right]$.
In accordance with the Vicsek model, we use the same settings as in
Ref. \cite{vi95}, i.e. $r=1$, $v=0.03$ and $N=300$, and employ the
absolute value of the average normalized velocity
$v_{a}=|\sum_{i=1}^{N}\mathbf{v}_{i}|/(Nv)$ as the performance
index. The velocity $v_a$ is approximately zero if the direction of
motion of the individual particles is distributed randomly, while
for the coherently moving phase (with ordered direction of
velocities) $v_a\simeq 1$. Note that the linear size $L$ of a square
shaped cell determines the density $\rho=N/L^{2}$, and in all the
simulations we use 1000 runs and $M=500$ running steps for each run.

For partial communication, only some of the particles will broadcast
its position and velocity at each time step. The communication cost
$p$ is measured by the average number of broadcasting particles over
the total number of particles at each time step. To investigate
partial communication and find a highly efficient manner, we compare
three communication manners, namely, {\it random, continuous} and
{\it supervised communication}. The random communication manner
(resp. the continuous communication manner) demands that each
particle send $p\cdot M$ messages reporting its position and
velocity randomly (resp. continuously with randomly selected
beginning step). The supervised manner is an intelligent one, in
which each particle calculates the average direction of its
broadcasting neighborhood at each step. When the angular difference
between its and neighboring directions surpasses an angular
threshold $\theta_{t}$, it will broadcast its position and velocity
to its neighbors. Obviously, the communication cost increases with
decreasing $\theta_{t}$.

\begin{figure}[htp]
\centering
\begin{tabular}{cc}
\hspace*{-0.3cm}
\resizebox{4.3cm}{!}{\includegraphics[width=11.2cm]{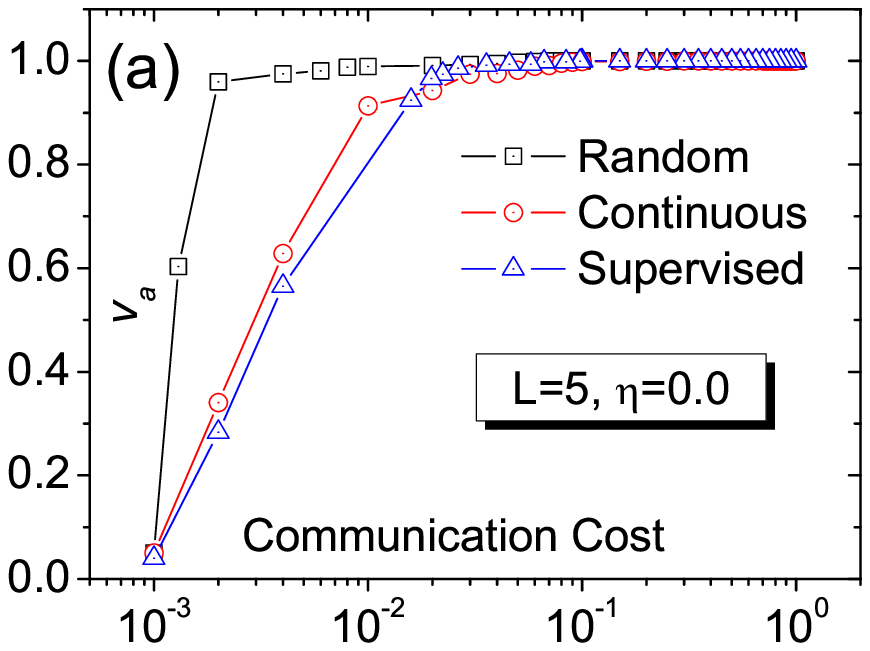}} &
\resizebox{4.3cm}{!}{\includegraphics[width=11.2cm]{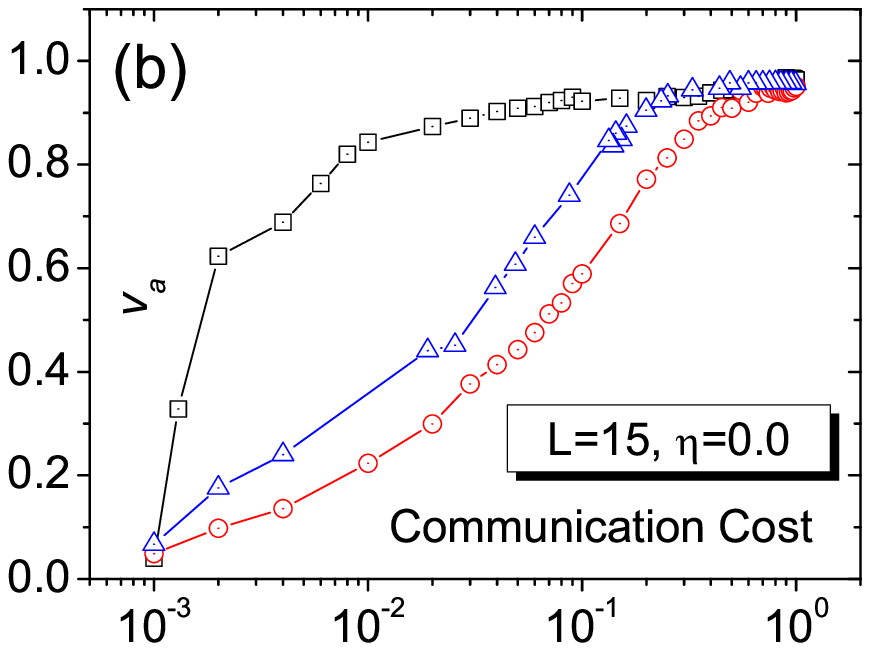}}
 \\
 \hspace*{-0.3cm}
\resizebox{4.3cm}{!}{\includegraphics[width=11.2cm]{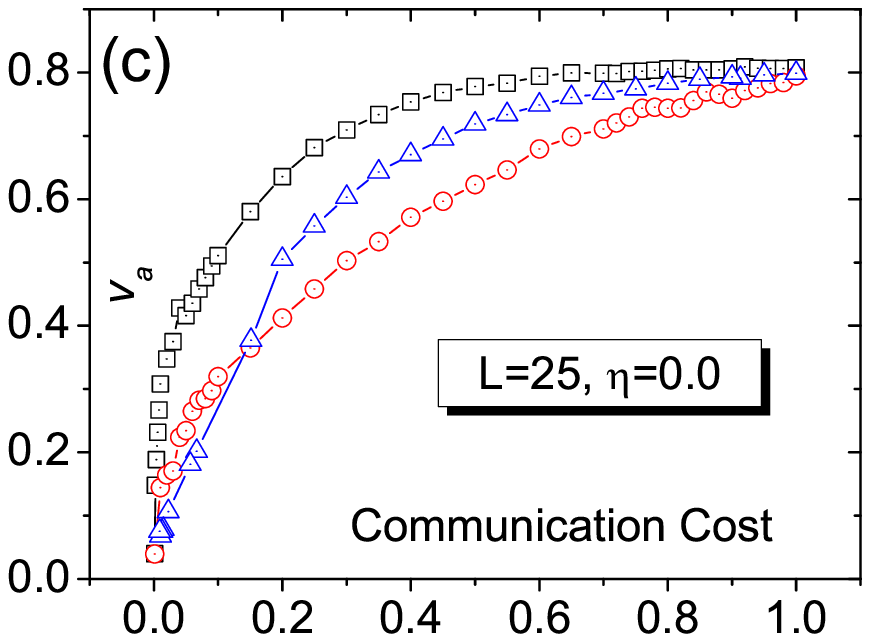}} &
\resizebox{4.3cm}{!}{\includegraphics[width=11.2cm]{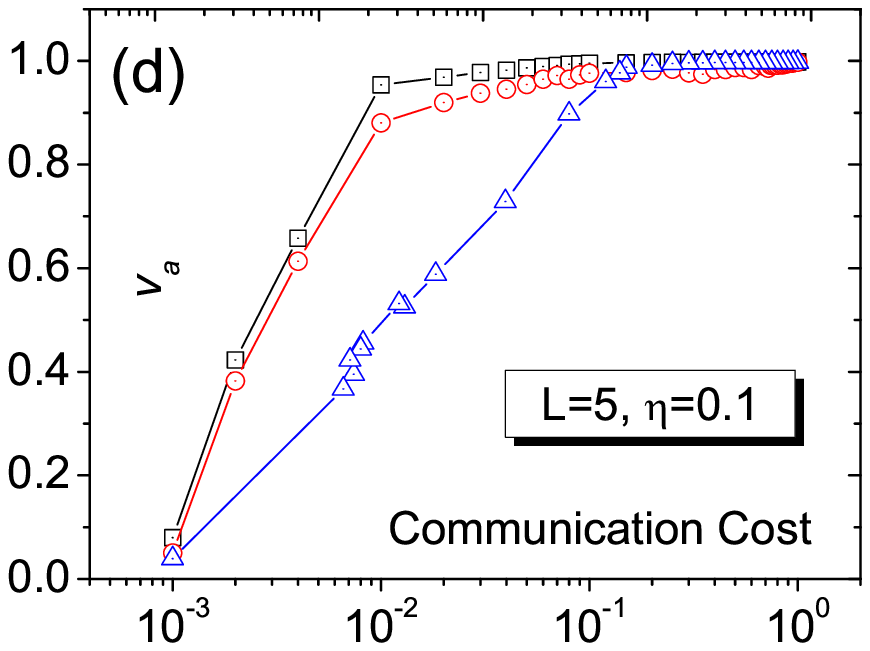}}
 \\
 \hspace*{-0.3cm}
\resizebox{4.3cm}{!}{\includegraphics[width=11.2cm]{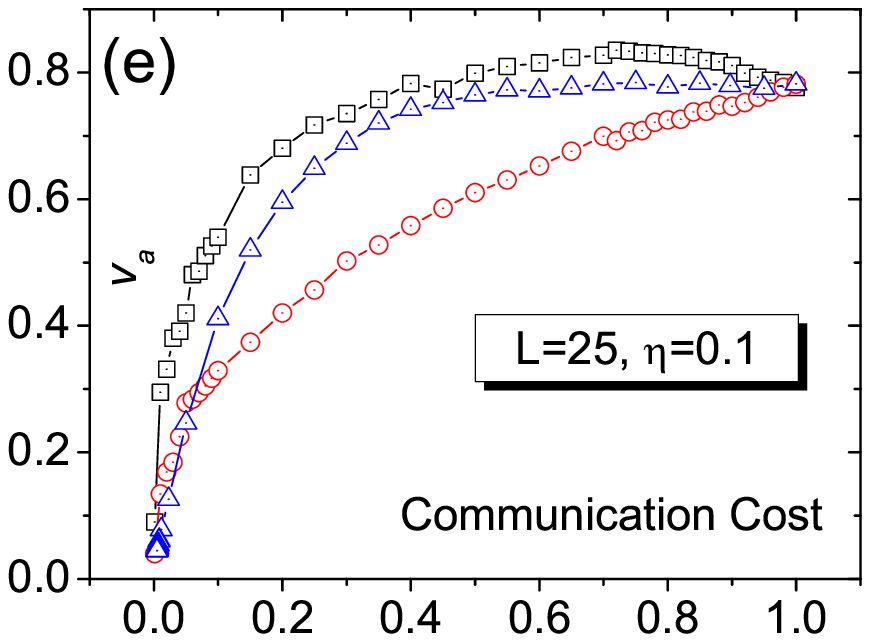}} &
\resizebox{4.3cm}{!}{\includegraphics[width=11.2cm]{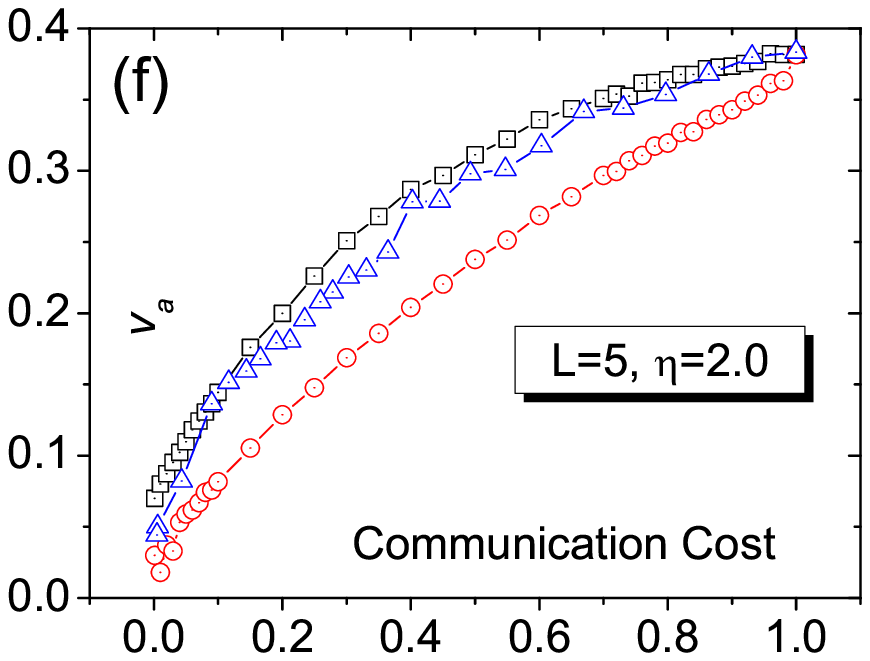}}
\end{tabular}
\caption{(color online) Synchronized performance index $v_a$ with
respect to communication cost $p$ for random (squares), continuous
(circles), and supervised (triangles) manners.}
 \label{fig: performance comparison}
\end{figure}

First, these three protocols are compared for a high density
particle group ($L=5$) without noise (i.e. $\Delta\theta=0$ in
Eq.(\ref{Eq Vicsek model angle})) in Fig.~\ref{fig: performance
comparison}(a). The first attractive characteristic of the random
manner is that satisfactory $v_{a}$ can be yielded with sharply
reduced $p$ (e.g. no less than 95$\%$ of $v_{a}$ of the complete
communication for $p\approx 2\%$). To further investigate the
influence of the density $\rho$ on the performance $v_{a}$, we have
done simulations for a medium density case ($L=15$) and a low
density case ($L=25$) (see Fig.~\ref{fig: performance comparison}(b)
and Fig.~\ref{fig: performance comparison}(c), respectively).
Comparing the performances for the random communication alone across
Fig.~\ref{fig: performance comparison}(a)--(c), one can observe
that, when the density $\rho$ decreases, to achieve the same
satisfactory $v_{a}$  a higher communication cost $p$ is required.
Similar conclusions can also be drawn for the continuous and
supervised protocols. Next, we investigate the effect of external
noise by adding low and high noises in Eq. (\ref{Eq Vicsek model
angle}). As shown in Fig.~\ref{fig: performance comparison}(d)--(f),
the noise has a more intensive effect on the performance than
density since the influence of the noise is more direct. It can be
seen that the random manner is the best among the three
communication manners.

The physical reason of achieving synchronized performances via very
low communication cost may be: due to the high density (e.g. $L=5$)
, on average there are enough particles inside the radius $r$ of
each particle, and the combination of the sparse messages of each
one of the plentiful neighbors constitutes an abundant information
flow which can guide each particle to the right direction. If the
density of particles is decreased (e.g. $L=15,25$), $p$ should be
increased to compensate for the deficiency of the neighboring
guidance information. As to why the random protocol is the best one
among the three, the explanation may be that communication manners
distributing messages more evenly are more efficient due to their
reduced redundancy. In detail, in the continuous manner, when one
message sent by a particle can guide its neighbors along the right
direction, its succeeding messages do little to help the group
performance. In this sense, these subsequent messages become
redundant, and the efficiency is thus decreased substantially. On
the other hand, since it is very natural to have a lurking suspicion
that a more intelligent communication protocol implies better
performance, it is surprising that the random manner is superior to
the supervised manner. This fact should also be accredited to the
communication redundancy. Statistical simulation shows that when a
particle sends a message to avoid deviation, its neighbors are apt
to send messages simultaneously. Thus it is of high probability that
almost all the particles send messages at the same time. This
supposition is supported by the simulations which show that the
messages mostly aggregate at the very beginning of the whole
procedure. As a result, communication redundancy is inevitable. In
brief, it is reasonable to deduce that the best manner is to
distribute the communication messages as evenly as possible along
the whole dynamic process, since it minimizes the communication
redundancy.
\begin{figure}[htp]
\centering
\begin{tabular}{cc}
\resizebox{4.1cm}{!}{\includegraphics[width=11.2cm]{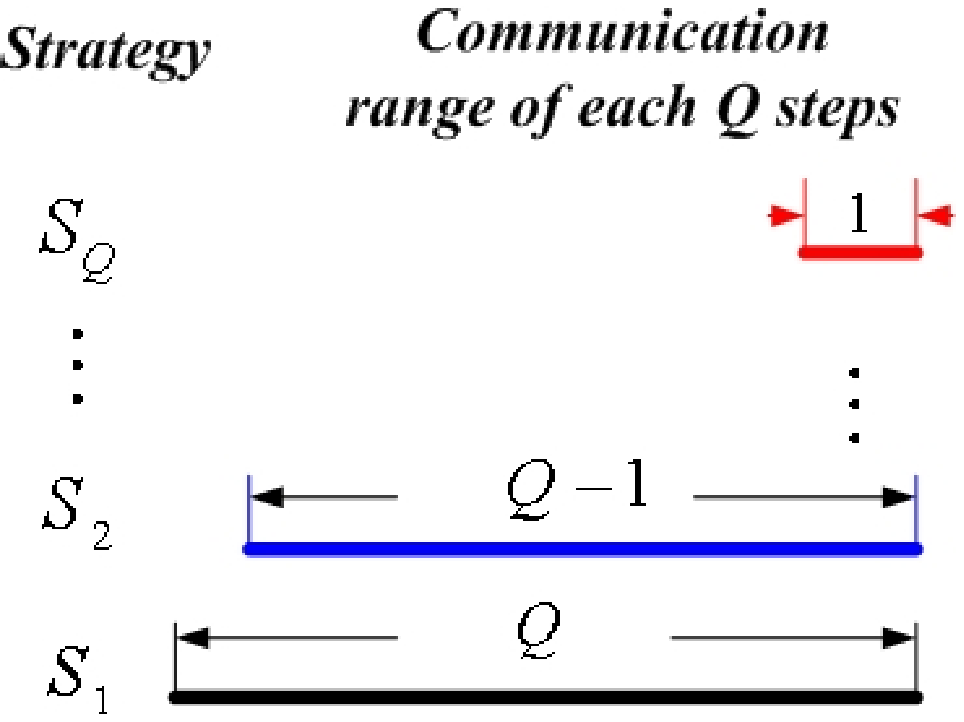}}
&
\resizebox{4.5cm}{!}{\includegraphics[width=11.2cm]{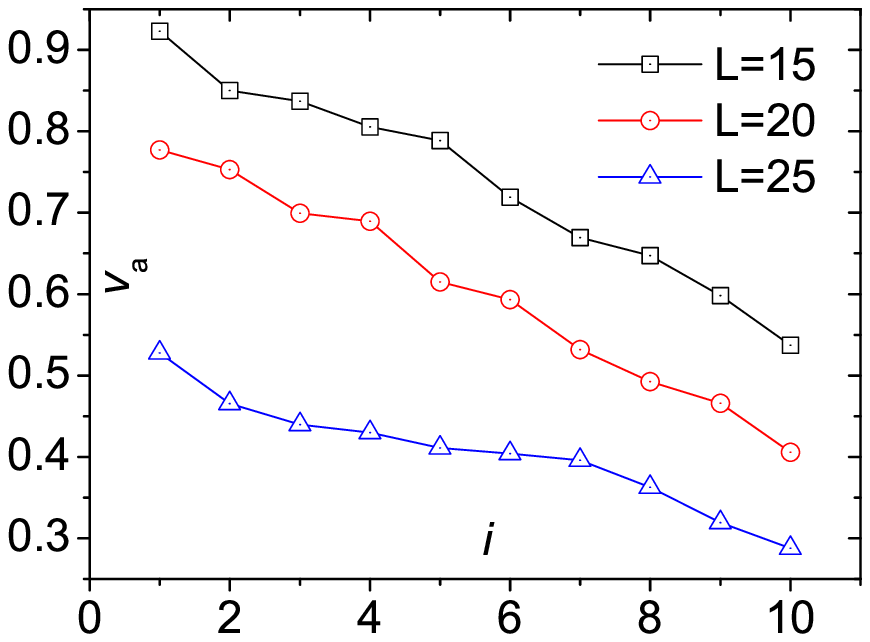}}
 \\
{\scriptsize (a) } &  {\scriptsize (b) } \\
\end{tabular}
\caption{(color online) (a) Illustration of different communication
strategies, and (b) performance of strategy $S_i$
($i=1,2,\cdots,Q$). Here we set $Q=10$. }
 \label{fig: redundancy}
\end{figure}

We propose a toy model to demonstrate that more evenly distributed
communication leads to better performance. As shown in
Fig.~\ref{fig: redundancy}(a), with a give integer $Q$, for strategy
$S_i$ ($1\leq i \leq Q$), each particle sends its messages with
probability $1/(Q-i+1)$ if the current time step $t$ satisfies the
inequality $t~mod~Q \geq i-1$. It is obvious that the communication
cost $p$ equals $1/Q$ for each strategy $S_i$ ($i=1,2,\ldots,Q$) and
the communication redundancy increases with increasing $i$. From the
performance comparison in Fig.~\ref{fig: redundancy}(b), we can
observe that $v_a$ decreases with increasing communication
redundancy.

Furthermore, a surprising phenomenon is observed that, in the case
of the random manner with noise (e.g. $L=25,\eta=0.1$ in
Fig.~\ref{fig: performance comparison}(e)), there exist an abnormal
region where moderately reducing $p$  might even increase $v_{a}$.
Thus, for the groups working in this region, each particle can use
much less communication power to gain even better performance. The
applaudable physical rule behind this astonishing phenomenon may be:
in some suitable areas of the density-noise space, the influence of
noise defeats the counterpart of the neighboring communications but
it has not yet reached the extent of totally disordering the system
dynamics, thus in some range of $p$ (e.g. $0.5\leq p \leq 1$) more
communication means propagating more errors. Consequently, partial
communication outperforms complete communication. This rule can be
very useful in plentiful industrial applications, since more
benefits can be achieved with less communication cost in some
working conditions. However, note that this phenomenon is only found
in the random communication. As to the continuous and supervised
manners, their communication redundancy is too much to arouse this
abnormal phenomenon.

\begin{figure}[htb]
  \centering \leavevmode
    \resizebox{6cm}{!}{\includegraphics[width=11.2cm]{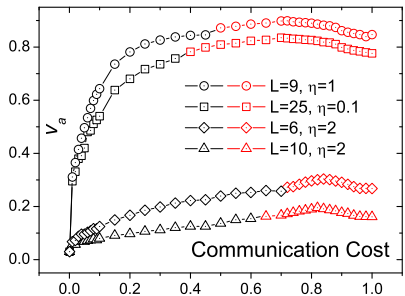}}
 \caption{(color online) Abnormal cases of random manner, red points denote abnormal phenomena.}
 \label{fig:abnormal_cases}
 \end{figure}

\begin{figure}[htb]
  \centering \leavevmode
    \resizebox{6cm}{!}{\includegraphics[width=11.2cm]
  {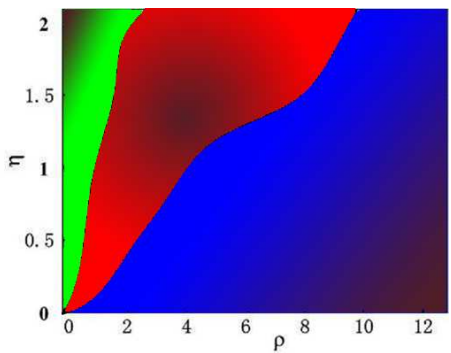}}
 \caption{(color online) Phase diagram: Abnormal (red), normal (blue) and disordered (green) regions.}
\label{fig:abnormal_region}
 \end{figure}

To illustrate the abnormal phenomenon of the random manner more
vividly, first we provide several typical abnormal cases
($L=25,\eta=0.1$; $L=6,\eta=2$; $L=10,\eta=2$ and $L=9,\eta=1$) and
mark their corresponding abnormal values by red points in
Fig.~\ref{fig:abnormal_cases}. We sketch the diagram in
Fig.~\ref{fig:abnormal_region} where the density-noise space is
divided into three regions, namely abnormal, normal and disordered
regions denoted by red, blue and green colors, respectively. Here,
the disordered region represents the density-noise combinations with
which the performance $v_{a}$ remains at a very low random value no
matter what $p$ is. Furthermore, the intensity of the color
represents the likelihood of the occurrence of each phenomenon. For
instance, the very inner part of the abnormal region is marked by
darker red color than the boundary, which means that in this central
part the abnormal phenomenon is more likely to occur. The
reasonableness of these regions can be validated by some simple
arguments as follows. There is no abnormal phenomenon in noise-free
cases, thus the line of $\eta=0$ always belongs to normal region;
for $\rho=\eta=0$, there is no particle, therefore the origin point
is the only intersection of the three regions. More importantly, in
the abnormal region (see red part of
Fig.~\ref{fig:abnormal_region}), the intensity of the noise has been
increased to defeat the influence of the neighboring communication.
However, if the noise is enhanced too quickly, then the system will
enter the disordered state (see the green part of
Fig.~\ref{fig:abnormal_region}). Using such density-noise space
diagrams, one can tell whether the current working condition of the
network is in the abnormal region or not. If so, one can estimate
how much communication energy can be saved to yield better
performance than complete communication. In this sense, the
discovery of such abnormal regions will be valuable in abundant
industrial applications.

In summary, we have numerically analyzed the collective dynamics of
self-propelled particle groups via partial communication and found
that: (i) Ordered state performance can be achieved with fairly low
communication cost. When the density is high, for noise-free or
low-noise cases, just a very small proportion of communication can
produce satisfactory performances; in other words, almost no benefit
can be gained by increasing the communication cost when $p$ exceeds
a very small value. (ii) There exist an abnormal region in the
density-noise space, in which moderately reducing the communication
cost can even improve the performance. (iii) More evenly distributed
communication is superior.

To verify the universality of these conclusions, we have also
applied the rule of partial communication to another two popular
models of self-propelled particles, the A/R model \cite{ga03} and
the Moreau model \cite{mo04}. The corresponding results also
strongly suggest that complete communication is not always optimal
when taking into consideration both the performance index and
communication cost.
 For natural science, the contribution of this
work is to explain why the particles of biological
flocks/swarms/schools like firefly and deep-sea fish groups do not
send their messages to others all along but just now and then during
the whole dynamic process. From the industrial application point of
view, the value of this work is two-fold. If the current working
condition is in the normal region, then the communication energy or
cost can be reduced very sharply at the cost of a tiny decrease of
synchronization performance, while in the abnormal region, the
minimum communication energy can be estimated to gain the maximum
benefit which is larger than the counterpart of complete
communication. This work is a first attempt aiming at achieving
satisfactory ordered state of a self-propelled particle group via
low communication cost, and we believe that it will enlighten the
readers on this interesting subject.


\end{document}